# Multiple-inheritance hazards in dependently-typed algebraic hierarchies

Eric Wieser[1[0000−0003−0412−4978]]

Cambridge University Engineering Department, efw27@cam.ac.uk

**Abstract.** Abstract algebra provides a large hierarchy of properties that a collection of objects can satisfy, such as forming an abelian group or a semiring. These classifications can arranged into a broad and typically acyclic directed graph. This graph perspective encodes naturally in the typeclass system of theorem provers such as Lean, where nodes can be represented as structures (or records) containing the requisite axioms. This design inevitably needs some form of multiple inheritance; a ring is both a semiring and an abelian group.

In the presence of dependently-typed typeclasses that themselves consume typeclasses as type-parameters, such as a vector space typeclass which assumes the presence of an existing additive structure, the implementation details of structure multiple inheritance matter. The type of the outer typeclass is influenced by the path taken to resolve the typeclasses it consumes. Unless all possible paths are considered judgmentally equal, this is a recipe for disaster.

This paper provides a concrete explanation of how these situations arise (reduced from real examples in **mathlib**), compares implementation approaches for multiple inheritance by whether judgmental equality is preserved, and outlines solutions (notably: kernel support for $\eta$-reduction of structures) to the problems discovered.

## 1 Introduction

It becomes clear very early in the development of mathematical libraries that a generalization over algebraic properties is essential; as soon as we are able to speak about $\mathbb{N}$ and $\mathbb{Z}$, we will want to have available that $a + b = b + a$ whether $a, b : \mathbb{N}$ or $a, b : \mathbb{Z}$, and it would be strongly preferable that we can refer to this property by a single name.

The generalization we seek is of course well-studied as the field of abstract algebra, and the commutativity property above can be phrased as "$\mathbb{N}$ and $\mathbb{Z}$ are both semirings", or using language more precise to the specific property we care about "$\mathbb{N}$ and $\mathbb{Z}$ are both abelian monoids". At least when considering only those which operate on a single carrier type, algebraic structures can be connected into a directed graph; all rings are semirings and abelian groups, so we can draw a pair of edges from "ring" to "semiring" and "abelian group". An illustration of the depth and breadth of such a graph can be seen in [19, fig. 1], while a reduced example that we will use in this paper can be seen in fig. 1.

Encoding this directed graph into the machinery of a particular theorem prover can be done in multiple ways, which are outlined in [3, §1] and presented with example code across a variety of languages in [6, fig. 1]. This paper focuses on the typeclass approach used by `mathlib` [19] in the Lean 3 theorem prover [14]; though the observations generalize to other implementations in dependent type theory built upon "structure" types.

In this approach, the graph is pruned to be acyclic, and then a typeclass is created for each node carrying its operators (data fields) and the properties they satisfy (proof fields). The edges correspond to functions converting from stronger structures to weaker structures, each registered as a typeclass instance. This encodes naturally in "record" or "structure" types with multiple inheritance, where we can write down the desired edges declaratively in the form of a list of base structures, and have the language generate the necessary "forgetful" instances automatically. A simple example of this can be found in [3, §4].

Unfortunately, the devil is in the details; in Lean, Coq, Agda, and Isabelle, support for multiple inheritance is not part of the underlying type theory, so types that use multiple inheritance have to be translated by the elaborator into inductive types that do not. There are multiple ways to perform such a translation, and the choice is not inconsequential.

In section 2 we outline two such approaches, and show how they can each be used to construct a much-reduced version of `mathlib`'s abstract algebra library. Section 3 introduces a more complex use of a typeclass from `mathlib`, and demonstrates how in the absence of special kernel support for $\eta$-reduction on structure types, its design is incompatible with "nested" approach to structures. Section 4 outlines some workarounds that permit the "nested" approach to be used even in the absence of this support. Section 5 explains how the problem is not unique to typeclass-based approaches.

The problems explored here are far from hypothetical; the migration of `mathlib` from Lean 3 to Lean 4 [15] forces a switch from the approach in section 2.1 to that in section 2.2, which has presented a significant stumbling block [5].

## 2   Types of structure inheritance

Lean 3 supports two types of structure inheritance: the default "new style", which we will refer to as "nested", and does not support multiple inheritance; and the legacy "old style" (enabled with `set_option old_structure_cmd true`) which we will refer to as "flat", and *does* support multiple inheritance. Lean 4 (as a language) does away with the "flat" mode, but extends the "nested" mode to support multiple inheritance.

To compare these approaches, this section demonstrates how to build the miniature algebraic hierarchy shown in fig. 1. If we permit ourselves to use the

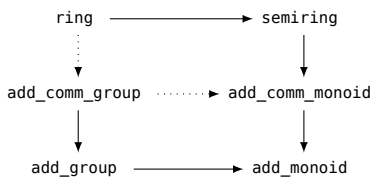

Fig. 1: A hierarchy of algebraic typeclasses, where arrows indicate a stronger typeclass implying a weaker typeclass. Dotted arrows correspond to the "non-preferred" typeclass paths which are relevant to section 2.2.

```
class add_monoid (α : Type) :=
(zero : α) (add : α → α → α)
```

```
class add_comm_monoid (α : Type) extends add_monoid α
```

```
class semiring (α : Type) extends add_comm_monoid α :=
(one : α) (mul : α → α → α)
```

```
class add_group (α : Type) extends add_monoid α :=
(neg : α → α)
```

```
class add_comm_group (α : Type) extends add_group α, add_comm_monoid α
```

```
class ring (α : Type) extends semiring α, add_comm_group α
```

Listing 1: The hierarchy in fig. 1 described using `extends` clauses.

builtin language support for multiple inheritance, we could write this as in listing 1. As they are not going to be relevant to the discussion in this paper, the proof fields such as `one_mul : ∀ a : α, mul one a = a` have all been omitted.

To avoid this paper being about a specific implementation of inheritance in a specific version of Lean, we will avoid the `extends` keyword, instead emulating it via different possible encodings of inheritance into regular structures. For simplicity this paper is largely presented as about Lean, but the supplemental repository referenced in section 7 demonstrates how the Lean 3 samples presented here can be replicated in Coq[1] and in Lean 4[2].

## 2.1 Flat structures

The "flat" approach to structure inheritance is to copy all of the fields from the base classes into the derived class. If multiple base classes share a field of the same name, then these fields are merged[3]. The forgetful instances are then implemented by unpacking all the relevant fields of the derived class and passing them to each base class constructor (which in Lean can be written as `{ ..derived }`).

---

[1] Albeit somewhat non-idiomatically.

[2] At least, in old versions without pertinent fixes!

[3] Unless they are of different types, which raises an error.

This can be seen for the toy example from listing 1 in listing 2a; `ring` extends both `semiring` and `add_comm_group`, so inherits the union of the four fields of `semiring` (`zero`, `add`, `one`, `mul`) and the three fields of `add_comm_group` (`zero`, `add`, `neg`). The `ring.to_semiring` and `ring.to_add_comm_group` instances generate constructor applications that reassemble the corresponding fields.

This approach is straightforward to implement in a theorem prover, and is the one used (via **set_option** `old_structure_cmd true`) in the majority of mathlib's algebraic hierarchy in Lean 3. A downside to this approach is that it can produce more work for unification (leading to poor performance) in long inheritance chains [3, §10].

(a) The flat approach (section 2.1), copying base fields to derived classes.

(b) The nested approach (section 2.2), inserting the first parent as a field and copying the remaining fields.

Listing 2: Two approaches to implementing inheritance, by elaborating the **extends** clauses in listing 1 as the highlighted lines.

## 2.2 Nested structures

A naïve approach to multiple inheritance for `ring` would be simply to create a structure containing a `to_semiring` field and a `to_add_comm_group` field. The problem with this approach is that the resulting structure contains two separate `add` fields. Compatibility of these fields could in principle be enforced with a proof field along the lines of `add_ok : to_semiring.add = to_add_comm_group.add`, but this makes the API very unpleasant to use as the user now has to rewrite between all the different copies of `add`.

The way to modify this approach to avoid this pitfall is to add a field for each base class that doesn't overlap with any previous base classes, otherwise fall back to the "flat" approach and add the non-overlapping fields directly. We call these non-overlapping base-classes "preferred" instances, as the projections for these fields can be registered directly with the typeclass system using `attribute [instance] derived.to_base`. What remains are the "non-preferred" instances, which can be constructed in a similar way to what was done in section 2.1, though with somewhat messier expressions. Note that unlike section 2.1, this approach is influenced by the order of the base classes.

This can be seen in listing 2b; `ring` contains a `to_semiring` field for its first base class, but `add_comm_group` would overlap so its remaining non-overlapping field (`neg`) is added separately. The "preferred" `ring.to_semiring` projection is then registered with the typeclass system, while the "non-preferred" `ring.to_add_comm_group` is painstakingly assembled piece-by-piece. To encourage Lean to avoid the "non-preferred" instance, we give it a low priority of 100 (the default is 1000).

This approach is more complicated to implement (and indeed, was not implemented in Lean until Lean 4), but can have performance advantages for unification as the "preferred" instance paths do not introduce a constructor application.

The result of listing 2b is that the graph in fig. 1 is imbued with an asymmetry; the dotted edges are provided by "non-preferred" instances. These edges can be chosen on any spanning tree[4] of the overall graph, and indeed can be optimized to fall on the paths most used by the library [11].

For the purpose of this paper, the opposite is true; their placement has been pessimized to deliberately cause a failure, which we shall see in section 3.2!

# 3 Typeclasses depending on typeclasses

In section 2, we concerned ourselves with the typical examples of typeclasses which depend on a single type. In Lean, it is possible for typeclasses to depend not only on multiple types, but on typeclasses that constrain those types. A simple typeclass of this form is `module R M`, which is used to declare that given a semiring $R$ and an abelian monoid $M$, there is an $R$-module structure on $M$. A more complete explanation of this typeclass can be found in [20] and [3, §5]. For the purpose of this paper, we can imagine the simpler definition as follows:

---

[4] In general this is a spanning diamond-free directed acyclic graph, but for this paper it suffices to consider a tree.

```
class module (R M : Type) [semiring R] [add_comm_monoid M] :=
(smul : R → M → M)
-- (one_smul : ∀ (x : M), smul 1 x = x)
-- (mul_smul : ∀ (r s : R) (x : M), smul (r * s) x = smul r (smul s x))
-- (add_smul : ∀ (r s : R) (x : M), smul (r + s) x = smul r x + smul s x)
-- (zero_smul : ∀ (x : M), smul 0 x = 0)
```

Here, the proof fields within the typeclass depend on the operators imbued upon the types $R$ and $M$. Just as in section 2, we shall ignore these proof fields as they are not relevant to the discussion other than providing motivation for the `[semiring R] [add_comm_monoid M]` parameters.

### 3.1  Equality of typeclass arguments

A natural use of this typeclass is to record the fact that any semiring is a module over itself, where the scalar action `smul` is just multiplication [20, §2.1]. This can be written in Lean as

```
instance semiring.to_module (R) [iS : semiring R] : module R R :=
{ smul := semiring.mul }
```

The type of this instance is misleading; while a human reader could be forgiven for assuming that the type is just `module R R`, to Lean the type is

```
@module R R iS (@semiring.to_add_comm_monoid R iS)
```

where `@` is syntax to tell Lean that even the automatically-populated typeclass arguments should be spelled out explicitly[5]. The expressions for these implicit arguments are visualized graphically in fig. 2a.

Lean can now tell us that a *ring* is a module over itself, as after all every ring is also a semiring. We can ask this question with:

```
example (R) [iR : ring R] : module R R := by apply_instance
```

Once again, the type is misleading; the true type can be seen in fig. 2b. Comparing the types for fig. 2a and fig. 2b, we see that the former unifies with the latter by setting `iS = @ring.to_semiring R iR`; for this reason, Lean finds our instance as `@semiring.to_module R (@ring.to_semiring R iR)`.

### 3.2  Inequality of typeclass arguments

Let's imagine now that we want to write a lemma that applies to a module over a ring (as opposed to a semi-module over a semiring), and states that $(-r)m = -(rm)$. We write this as[6]

```
lemma neg_smul {R M} [ring R] [add_comm_group M] [module R M] (r : R) (m : M) :
  module.smul (add_group.neg r) m = add_group.neg (module.smul r m) := sorry
```

---

[5] This style of display can be enabled with `set_option pp.implicit true` in Lean 3 and `set_option pp.explicit true` in Lean 4.

[6] Omitting the usual · and ⋆ notation to keep listing 2 short.

To complete our setup, let's check that this lemma applies to the $R$-module structure on $R$:

```
example {R} [iR : ring R] (r : R) (r' : R) :
  module.smul (add_group.neg r) r' = add_group.neg (module.smul r r') :=
neg_smul r r'
```

If we use the "flat" design in listing 2a, then this continues to work as expected. The same is not true of the "nested" design in listing 2b, which fails to synthesize type class instance for

```
@module R R (@ring.to_semiring R iR)
  (@add_comm_group.to_add_comm_monoid R (@ring.to_add_comm_group R iR))
```

which is shown graphically in fig. 2c. The `neg_smul` lemma is an example of how typeclass resolution can be steered through a specific node of the graph in fig. 1.

In Lean 3, the reason this fails is nothing to do with typeclass search; the problem is that the type in fig. 2c is not equal to type in fig. 2b, due to the implicit `add_comm_monoid M` arguments (shown in red) not being considered equal. Considerations of equality between the red paths in figs. 2b and 2c are often referred to as a "typeclass diamonds" due to the shape they form when overlaid; though this is a rather more subtle diamond problem that the ones described in [20, §5] and [2, §3.1] as it is caused by code that would normally be invisible to the user.

To mathematicians, this diagram obviously commutes; weakening a ring to an abelian monoid via a semiring is the same as doing so via an abelian group. But Lean doesn't care about "obviously": when determining equality of types, it's not enough for them to just be *provably* the same; they need to be *definitionally* (sometimes called judgmentally) so. A proof of `rfl` can be used to determine if two terms are judgmentally equal; under listing 2b, we get an error confirming they are not:

```
example (R) [iR : ring R] :
  (@semiring.to_add_comm_monoid R (@ring.to_semiring R iR)) =
  (@add_comm_group.to_add_comm_monoid R (@ring.to_add_comm_group R iR)) :=
rfl -- fails in Lean 3 with listing 2b
```

### 3.3  Impact of the inheritance strategy

The `rfl` in section 3.2 that fails under listing 2b but not listing 2a tells us that the nested inheritance is certainly to blame here. The underlying cause is the difference between the "preferred" and "non-preferred" paths.

The "non-preferred" edges in listing 2b are implemented directly as a constructor application via the `{ }` syntax; so by virtue of following "non-preferred" edges, the red path in fig. 2c unfolds to an application of the `add_comm_monoid` constructor. The "preferred" edges correspond to a projection; unless applied to something that unifies against a constructor, these operations themselves do not unify against a constructor. As the red path in fig. 2b consists of only "preferred" edges, it only unifies with this `add_comm_monoid` constructor if `iR` unifies with a `ring` constructor.

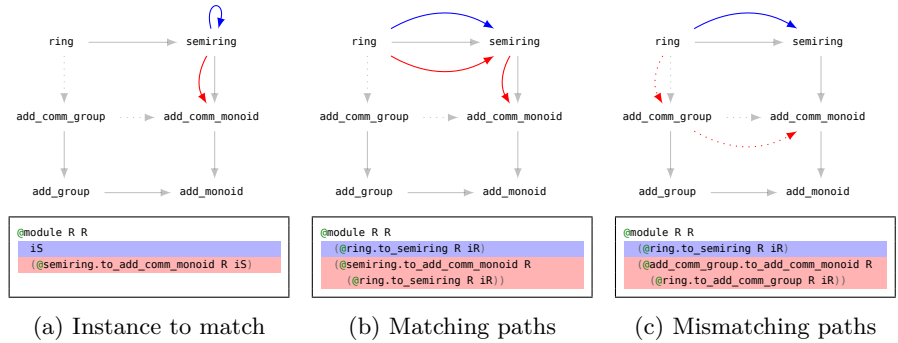

(a) Instance to match    (b) Matching paths    (c) Mismatching paths

Fig. 2: Paths taken through the graph in fig. 1 when filling the two implicit arguments of the type of `module R R`. Dotted lines again refer to "non-preferred" edges.

If `iR` is a concrete instance such as `instance int.ring : ring ℤ`, then it will almost certainly unify with a `ring` constructor, and the overall unification problem is solvable. However if `iR` is a free variable, it will only unify with a constructor in systems which support "η-reduction for structures". Lean 3 is not such a system, which makes unification impossible.

### 3.4 Other examples in `mathlib`

The `module` typeclass is far from the only typeclass in `mathlib` that follows the pattern introduced in section 3; some others typeclasses (all of which fall afoul of the issue in section 3.2) include

- `algebra` (`R A : Type`) `[comm_semiring R]` `[semiring A]`, indicating that $A$ is an $R$-algebras.
- `star_ring` (`R : Type`) `[non_unital_semiring R]`, indicating that there is a ⋆ operator compatible with the existing ring structure on $R$.
- `cstar_ring` (`R : Type`) `[non_unital_normed_ring R]` `[star_ring R]`, indicating that the existing norm, ⋆, and ring structure are suitable to declare $R$ a $C^*$-ring.

Like the `module` example, the design of the first of these is brought on by a need to work with two separate carrier types, and the need to avoid "dangerous instances" [3, §5.1].

The other two can be described as "mixin" typeclasses, and are motivated by a desire to avoid a combinatorial explosion of typeclass variations: an attempt at `star_ring` without mixins could easily end up needing all 16 variations of unital/non-unital commutative/non-commutative normed? star rings/fields. This motivation is largely a pragmatic one; the introduction of a tool like Coq's Hierarchy Builder [7] to `mathlib` would eliminate the cost of manually authoring such an explosion of typeclasses.

## 4  Mitigation strategies

### 4.1  Perform $\eta$-reduction of structures in the kernel

A key difference between the type theory of Lean 3 and Lean 4 is that Lean 4 adds a kernel reduction rule that $\eta$-reduces structures[7], which is precisely what we concluded we needed in section 3.3. The following example demonstrates what this means:

```
structure point := (x y : ℤ)

-- fails in Lean 3, succeeds in Lean 4
example (p : point) : p = { x := p.x, y := p.y } := rfl
```

In essence, any value from a `structure` type is considered judgmentally equal to its constructor applied to its projections.

This feature was motivated by various "convenience" definitional equalities (as requested by [8]), such as wanting `e.symm.symm = e` for an equivalence `e : α ≃ β`; but in a thankful coincidence happens to be precisely the tool needed to resolve the trap in section 3.2 that Lean 4 dropping support for "flat" structures would otherwise have ensnared us in. In particular, the Lean 4 version of the failing `example ... := rfl` above succeeds.

Until 2023-02-22, the structure $\eta$-reduction rule was disabled in Lean 4 during typeclass search; both due to performance concerns, and an absence of any evidence that it was necessary in the first place. As evidence mounted [5], a compromise was reached to unblock the Lean 4 version of `mathlib` that allowed it to be temporary enabled[8] in places where there was no other choice but taking the performance hit. After some unification performance improvements which are out of scope for this paper, this behavior was turned on globally on 2023-05-16 [9].

Lean 4 is not the only language to have taken an experimental approach to structural $\eta$; Coq supports it too, under the disabled-by-default `Primitive Projections` option. In contrast, Agda enables it by default for `inductive` types[9], but allows it to be disabled via `no-eta-equality`.

### 4.2  Use "flat" inheritance

The obvious approach to avoiding problems with "nested" inheritance is to simply not use it. Unfortunately, in the absence of elaborator support for translating a variation of listing 1 into listing 2a (such as in Lean 4) this would have to be done by hand, which can be rather tedious and error-prone.

There is however a trick; since the elaborator can translate listing 1 into listing 2b, we can construct a pathological graph such that all the edges we care

---

[7] Strictly speaking, it $\eta$-reduces `inductive` types with one constructor; `structures` are not native to the type theory of Lean, and instead just syntax for generating a suitable inductive type.

[8] Via `set_option synthInstance.etaExperiment true`.

[9] Some motivating discussion can be found in [1].

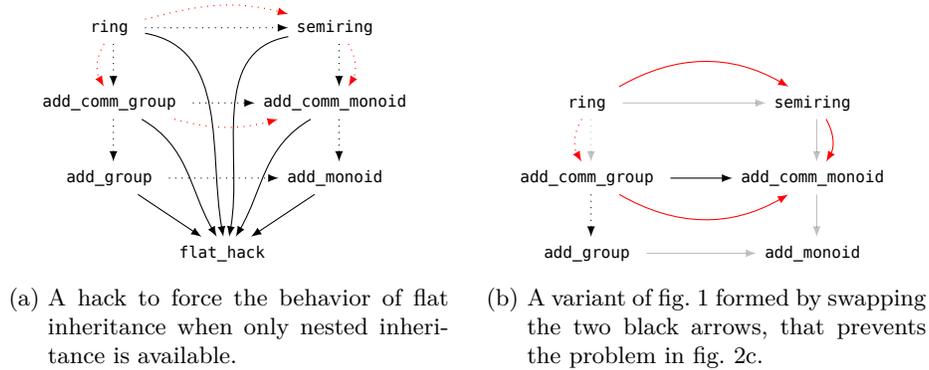

(a) A hack to force the behavior of flat inheritance when only nested inheritance is available.

(b) A variant of fig. 1 formed by swapping the two black arrows, that prevents the problem in fig. 2c.

Fig. 3: Alternate placements of the "preferred" spanning tree, with the diamond discussed in fig. 2 overlaid.

about are forced to be "non-preferred". We do this by adding an empty `flat_hack` structure as the first base class of every structure, which ensures that the base classes always overlap (due to the `to_flat_hack` field), and so the only "preferred" base class is the unused `to_flat_hack` projection. The spanning tree of "preferred" base classes across all such typeclasses is a star with `flat_hack` at its center, as shown in fig. 3a.

This forces all the typeclass resolution to go through the "non-preferred" paths, which behave identically to their "flat" counterparts by unfolding to a constructor application.

### 4.3   Carefully select "preferred" paths

In section 2.2, we mention that the choice of where to place the spanning tree of "preferred" paths could be optimized for performance. In light of section 3.2, we could instead attempt to optimize to ensure that the problematic diamonds never arise. Indeed, there are many arrangements of the "preferred" paths in fig. 1 that do not run into the *specific example* in fig. 2c, such as fig. 3b.

For our purposes, an adequate rule for why the red arrows of fig. 3 commute but the ones of fig. 2 do not is that the paths commute only if their last segments are either both "preferred" (as in fig. 3b) or both "non-preferred" (as in fig. 3a).

As discussed in [12] and visualized in fig. 4, it is not in general possible to choose a spanning tree for a set of 8 typeclasses arranged in a cube, while simultaneously making the pairs of paths around each face commute. This can be adapted into a working solution by inserting extra nodes in the style of section 4.2's `flat_hack` to force some additional paths to be "non-preferred", but this is far from an elegant solution.

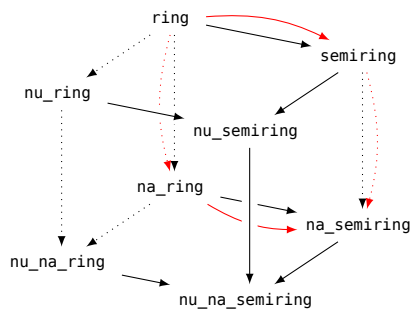

Fig. 4: An algebraic hierarchy where no spanning tree placement can ensure all squares commute, shown with one such inadequate spanning tree. The red paths highlight the one square that does not commute. `na` and `nu` are abbreviated from `mathlib`'s `non_unital` and `non_assoc`(iative).

### 4.4   Ban non-root structures in dependent arguments

The problem in section 3.2 is caused by a typeclass argument to a typeclass being inferable both via "preferred" and "non-preferred" routes. In section 4.2, this can be worked around by ensuring every path is maximally "non-preferred". An alternative is to ensure that every path is "preferred", by only accepting typeclass arguments that appear as roots of the spanning subgraph. This could look like

```
class module (R M : Type)
  [has_zero R] [has_add R] [has_one R] [has_mul R]
  [has_zero M] [has_add M] :=
(smul : R → M → M)
-- (one_smul : ∀ (x : M), smul 1 x = x)
-- (mul_smul : ∀ (r s : R) (x : M), smul (r * s) x = smul r (smul s x))
-- (add_smul : ∀ (r s : R) (x : M), smul (r + s) x = smul r x + smul s x)
-- (zero_smul : ∀ (x : M), smul 0 x = 0)
```

where each of the operators for $R$ and $M$ is taken as a separate typeclass argument.

This approach has two main downsides: it results in larger proof terms, because now it has 6 typeclass arguments instead of four, which have to be resolved all the way down to the smallest typeclass instead of stopping part-way along the graph; and it doesn't extend to cases where not just the data fields carrying the operators on the type arguments, but also the proof fields carrying their properties, are needed to define the fields of the dependent typeclass.

## 5   Implications for packed structures

Up until this point we have focused only on typeclasses, as these are (at the time of this paper) the idiomatic way to represent algebraic structure in Lean.

While Coq also supports typeclasses, and the previous examples can be faithfully reproduced in it, this is not the idiomatic way to do things in MathComp.

Instead, Coq's "Hierarchy builder" [7, §4] generates "packed" structures [10] with a field for the type itself, rather than consuming the type as a parameter. These structures are then ineligible for typeclass search, but can be located automatically via "canonical structures" (or as they are known in Lean, "unification hints") instead. These can in fact be built on top of the typeclasses from section 2.1 or section 2.2:

```
structure packed_semiring := (carrier : Type) [semiring carrier]
structure packed_add_comm_monoid := (carrier : Type) [add_comm_monoid carrier]
```

A naïve encoding of a module in this packed view would be:

```
structure packed_module :=
(R : packed_semiring) (M : packed_add_comm_monoid) [module R.carrier M.carrier]
```

As `packed_module` has no parameters and is therefore not dependently-typed, it cannot fall afoul of the problem in section 3.2.

Unfortunately, this encoding is effectively useless mathematically [18, §3]; we have no way to talk about two modules over the same ring without something involving equality of types and operators[10] like (`V W : packed_module`) (`hVW : V.R = W.R`); a much worse version of the duplicate `add` fields described at the start of section 2.2.

A more reasonable representation that avoids this problem is to only partially pack the structure, as

```
structure packed_module (R : packed_semiring) :=
(M : packed_add_comm_monoid M) [module R.carrier M.carrier]
```

which allows (`V W : packed_module R`). This is roughly analogous to the approach taken in Coq's MathComp [13] and in mathlib's category theory library.

While this representation avoids the *specific* problem in section 3.2 due to its type not depending on the `add_comm_monoid` path (the red arrows in fig. 2), it is nonetheless dependently-typed. This make it vulnerable to an analogous problem where the diamond is instead formed by the `semiring` path (the blue arrows in fig. 2) after adding two new `comm_semiring` and `comm_ring` nodes.

Fortunately for MathComp, the "Hierarchy builder" uses *flat* packed structures[11], and so avoids these issues for the same reason that flat typeclasses do in section 3.1.

---

[10] Or alternatively, by packing the ring and both modules into a single structure, as (`W : packed_module₂`) (`v : W.1`) (`w : W.2`). This is a viable approach for a module over two rings (as rarely are many rings needed), but doesn't scale for $n$ modules over the same ring.

[11] Presumably due to simplicity of implementation; there is no mention in [7] that using nested inheritance instead would have run into the issues described here.

## 6   Related work

While this work is of course directly related to the work of porting Lean 3's `mathlib` to Lean 4, the lessons here are transferable to Coq (where [7] seemingly correctly chose to use flat structures by coincidence) and Agda (which has adopted structure $\eta$-reduction globally due to other motivations [1]); even if only to provide further understanding of why the respective choices that have already been made in those systems are the correct ones. To the author's awareness, no previously demonstrated *algebraic* motivations have been given for $\eta$-reduction in the kernel. Some in-depth analysis of "coherence" in algebraic typeclass paths is provided by [17, definition 3.3] (another name for our comparison in fig. 2), but it does not provide an example to show *why* $\eta$-reduction specifically should be assumed.

The analysis in sections 3 and 4 is only relevant to systems that use dependent type theory, as concerns of equalities between the values of type parameters cannot arise in a language that does not permit those parameters in the first place. The Isabelle proof assistant which uses simple type theory is therefore immune to this class of problem; and at any rate [4, §5.4] advocates avoiding its record types entirely for algebraic structure, in favor of using locales.

Algebraic hierarchies certainly do not only exist in proof assistants; they are an essential part of computer algebra systems too. However, most computer algebra systems do not make use of dependent types [16, §1], with a notable exception being the Axiom Library Compiler, Aldor. Despite supporting dependent types, the type system of Aldor is too restrictive for sections 3 and 4 to be relevant. Aldor does not implement definitional equality of types (referred to as "value-equality" by [16, §2.4]), and so falls at a much earlier hurdle than the one in section 3; it does not consider `Vector(2+3)` and `Vector(5)` to be the same type [16, §2.3], meaning that even fig. 2b would be considered a mismatch, and *every* square in fig. 4 would not commute.

This work focuses on how a seemingly innocuous implementation detail can be crucial to ensuring the success of *existing* approaches to algebraic hierarchies in dependently-typed proof assistants. The broader analysis of these hierarchies, and possible alternative designs (for which computer algebra systems can provide inspiration), is left to [3, 6, 7, 18].

## 7   Conclusion

In this paper we have shown that for the "nested" approach to multiple inheritance to be viable in the context of dependently-typed typeclasses or packed structures, either we have to severely restrict how such inheritance is used (sections 4.2 to 4.4), or the kernel of the theorem prover must implement $\eta$-reduction for structures (section 4.1).

This scenario was a major stumbling block for `mathlib`'s transition from Lean 3 to Lean 4, as typeclasses of this form are used extensively in linear algebra. This paper provides a clear explanation of exactly what was going wrong, and a selection of various solutions that were considered before ultimately settling on the kernel change.

The code examples throughout this paper, along with translations into Lean 4 and Coq, and the version information needed to run them, can be found at `https://github.com/eric-wieser/lean-multiple-inheritance`.

**Acknowledgments** The author is grateful to: Gabriel Ebner, for campaigning for $\eta$-reduction support in Lean 4; Kazuhiko Sakaguchi, for providing insight into analogous situations in Coq; the anonymous referees, as well as Yaël Dillies and Filippo A. E. Nuccio, for valuable feedback on the manuscript; and the wider Lean community for collaboratively diagnosing [5] that the diamond problems discussed in section 3.2 existed. The author is funded by a scholarship from the Cambridge Trust.

# References

1. Abel, A.: On Extensions to Definitional Equality in Agda (Sep 2009), `https://www.cse.chalmers.se/~abela/talkAIM09.pdf`
2. Affeldt, R., Cohen, C., Kerjean, M., Mahboubi, A., Rouhling, D., Sakaguchi, K.: Competing Inheritance Paths in Dependent Type Theory: A Case Study in Functional Analysis. In: Peltier, N., Sofronie-Stokkermans, V. (eds.) Automated Reasoning, vol. 12167, pp. 3–20. Springer International Publishing, Cham (2020). `https://doi.org/10.1007/978-3-030-51054-1_1`, series Title: Lecture Notes in Computer Science
3. Baanen, A.: Use and abuse of instance parameters in the Lean mathematical library. In: ITP 2022. Haifa, Israel (May 2022), `http://arxiv.org/abs/2202.01629`
4. Ballarin, C.: Exploring the Structure of an Algebra Text with Locales. Journal of Automated Reasoning **64**(6), 1093–1121 (Aug 2020). `https://doi.org/10.1007/s10817-019-09537-9`
5. Buzzard, K.: leanprover/lean4#2074: typeclass inference failure (Jan 2023), `https://github.com/leanprover/lean4/issues/2074`
6. Carette, J., Farmer, W.M., Sharoda, Y.: Leveraging the Information Contained in Theory Presentations. In: Benzmüller, C., Miller, B. (eds.) Intelligent Computer Mathematics, vol. 12236, pp. 55–70. Springer International Publishing, Cham (2020). `https://doi.org/10.1007/978-3-030-53518-6_4`
7. Cohen, C., Sakaguchi, K., Tassi, E.: Hierarchy Builder: Algebraic hierarchies Made Easy in Coq with Elpi (System Description). In: Ariola, Z.M. (ed.) 5th International Conference on Formal Structures for Computation and Deduction (FSCD 2020). Leibniz International Proceedings in Informatics (LIPIcs), vol. 167, pp. 34:1–34:21. Schloss Dagstuhl–Leibniz-Zentrum für Informatik, Dagstuhl, Germany (2020). `https://doi.org/10.4230/LIPIcs.FSCD.2020.34`, iSSN: 1868-8969
8. Ebner, G.: leanprover/lean4#777: Definitional eta for structures (Nov 2021), `https://github.com/leanprover/lean4/issues/777`
9. Ebner, G.: leanprover/lean4#2210: Skip proof arguments during unification, and try structure eta last (May 2023), `https://github.com/leanprover/lean4/pull/2210`
10. Garillot, F., Gonthier, G., Mahboubi, A., Rideau, L.: Packaging Mathematical Structures. In: Berghofer, S., Nipkow, T., Urban, C., Wenzel, M. (eds.) Theorem

Proving in Higher Order Logics, vol. 5674, pp. 327–342. Springer Berlin Heidelberg, Berlin, Heidelberg (2009). `https://doi.org/10.1007/978-3-642-03359-9_23`, series Title: Lecture Notes in Computer Science

11. Gouëzel, S.: leanprover-community/mathlib4#3840: tweak priorities for linear algebra (May 2023), `https://github.com/leanprover-community/mathlib4/pull/3840`

12. Gouëzel, S.: #mathlib4 > Some observations on eta experiment (May 2023), `https://leanprover.zulipchat.com/#narrow/stream/287929-mathlib4/topic/Some.20observations.20on.20eta.20experiment/near/355336941`

13. Mahboubi, A., Tassi, E.: Mathematical Components (Sep 2022), `https://zenodo.org/record/7118596`, publisher: Zenodo Version Number: 1.0.2

14. Moura, L.d., Kong, S., Avigad, J., Doorn, F.v., Raumer, J.v.: The Lean Theorem Prover (System Description). In: Automated Deduction - CADE-25. pp. 378–388. Springer, Cham (Aug 2015). `https://doi.org/10.1007/978-3-319-21401-6_26`

15. Moura, L.d., Ullrich, S.: The Lean 4 Theorem Prover and Programming Language. In: Platzer, A., Sutcliffe, G. (eds.) Automated Deduction – CADE 28, vol. 12699, pp. 625–635. Springer International Publishing, Cham (2021). `https://doi.org/10.1007/978-3-030-79876-5_37`, series Title: Lecture Notes in Computer Science

16. Poll, E., Thompson, S.: Integrating Computer Algebra and Reasoning through the Type System of Aldor. In: Kirchner, H., Ringeissen, C. (eds.) Frontiers of Combining Systems. pp. 136–150. Springer Berlin Heidelberg, Berlin, Heidelberg (2000)

17. Sakaguchi, K.: Validating Mathematical Structures. In: Peltier, N., Sofronie-Stokkermans, V. (eds.) Automated Reasoning, vol. 12167, pp. 138–157. Springer International Publishing, Cham (2020). `https://doi.org/10.1007/978-3-030-51054-1_8`

18. Spitters, B., Van Der Weegen, E.: Type classes for mathematics in type theory. Mathematical Structures in Computer Science **21**(4), 795–825 (Aug 2011). `https://doi.org/10.1017/S0960129511000119`

19. The mathlib Community: The lean mathematical library. In: Proceedings of the 9th ACM SIGPLAN International Conference on Certified Programs and Proofs. pp. 367–381. ACM, New Orleans LA USA (Jan 2020). `https://doi.org/10.1145/3372885.3373824`

20. Wieser, E.: Scalar actions in Lean's mathlib. In: CICM 2021. Timisoara, Romania (Aug 2021), `http://arxiv.org/abs/2108.10700`